\documentclass[10pt,twocolumn,letterpaper]{article}

\usepackage{cvpr}              

\usepackage{graphicx}
\usepackage{amsmath}
\usepackage{amssymb}
\usepackage{booktabs}

\usepackage[pagebackref,breaklinks,colorlinks]{hyperref}

\usepackage[capitalize]{cleveref}
\crefname{section}{Sec.}{Secs.}
\Crefname{section}{Section}{Sections}
\Crefname{table}{Table}{Tables}
\crefname{table}{Tab.}{Tabs.}

\begin{document}

\title{Multimodal Political Bias Identification and Neutralization}

\author{Cedric Bernard\\
{\tt\small cedricb@vt.edu}
\and
Xavier Pleimling\\
{\tt\small xavierp7@vt.edu}
\and
Amun Kharel\\
{\tt\small akharel@vt.edu}
\and
Chase Vickery\\
{\tt\small cdvickery@vt.edu}
}

\maketitle

\begin{abstract}
Due to the presence of political echo chambers, it becomes imperative to detect and remove subjective bias and emotionally charged language from both the text and images of political articles.  However, prior work has focused on solely the text portion of the bias rather than both the text and image portions.  This is a problem because the images are just as powerful of a medium to communicate information as text is.  To that end, we present a model that leverages both text and image bias which consists of four different steps.  Image Text Alignment focuses on semantically aligning images based on their bias through CLIP models.  Image Bias Scoring determines the appropriate bias score of images via a ViT classifier.  Text De-Biasing focuses on detecting biased words and phrases and neutralizing them through BERT models.  These three steps all culminate to the final step of debiasing, which replaces the text and the image with neutralized or reduced counterparts, which for images is done by comparing the bias scores. The results so far indicate that this approach is promising, with the text debiasing strategy being able to identify many potential biased words and phrases, and the ViT model showcasing effective training. The semantic alignment model also is efficient.  However, more time, particularly in training, and resources are needed to obtain better results.  A human evaluation portion was also proposed to ensure semantic consistency of the newly generated text and images.
\end{abstract}

\section{Problem Statement}
In a study conducted in 2015, researchers analyzed approximately 150 million tweets from 3.8 million Twitter Users, each pertaining to political and nonpolitical issues. In this study, researchers observed that people who share the same ideologies when it comes to political topics exchange information with each other much more than those who share different ideologies \cite{barbera_tweeting_2015}. Through empirical observation, the Internet and social media are the primary news distributors today. Since we distribute news on a public platform, social media either acts as a public sphere that can host a variety of opinions and information or an echo chamber that strengthens views that it is built upon \cite{colleoni_echo_2014}. 

We postulate that the most significant contributor to the presence of political echo chambers in social media is the presence of subjective bias or emotionally charged language in the texts. In addition, propaganda is widespread in the news through cleverly crafted political images. For a healthy exchange of ideas between people of different political persuasions in social media, political images, and text should adhere to Dahlberg’s six essential qualities of the Public Sphere. The six qualities are namely: Reasoned Exchange of Problematic Validity Claims, Reflexivity, Ideal Role Taking, Sincerity, Formal Inclusion and Discursive Equality, and Autonomy from State and Corporate Power \cite{robertson_social_2018,dahlberg_habermasian_2004}. We need to cultivate these essential qualities in the domain of news and news distribution by debiasing images and texts in the news to make them objective and neutral regardless of their political standing.

In previous work on text debiasing in the context of news, researchers aimed to debias text by finding methods to make the text more neutral \cite{pryzant_automatically_2019}. Text debiasing was done by removing subjective bias using the method provided by \cite{recasens_linguistic_2013}: using Wikipedia’s Neutral Point of View (NPOV) policy. However, for debiasing political images, the work performed on it is very limited and not well explored. We will perform cross-modal alignment of textual images and news to solve the problem of debiasing political images. Then we will select the image with the least bias using the method in \cite{thomas_predicting_nodate}. In summary, our work attempts to learn an algorithm to minimize and neutralize political bias in news articles.

\section{Related Works}

\subsection{Debiasing}
Machine learning and deep learning models are trained on a large set of texts and images from the real world. These images and texts may contain gender, cultural, religious, political, and other social biases \cite{liang_towards_2020}. Several works have been proposed to address the problem of biases in real-world models. Some previous work was centered around debiasing the sentence level representations, which removes the religious, gender, racial, and cultural biases \cite{liang_towards_2020,manzini_black_2019}. \cite{liang_towards_2020} in particular analyzes the performance of debiasing on sentence-level downstream tasks such as sentiment analysis, linguistic acceptability, and natural language understanding.

For political bias detection, \cite{chen_analyzing_2020} uses over 6900 news articles with labels derived from a website to develop a neural model for bias assessment, and \cite{pryzant_automatically_2019} both analyzes subjective bias in text and neutralizes the subjective bias. There are two core problems with political news, which are namely subjective or biased language and news, and selective news reporting. \cite{pryzant_automatically_2019} is used to address the first issue of modifying subjective bias in a paper, giving an example where the paper converts the sentence “John McCain exposed as an unprincipled politician” to “John McCain described as an unprincipled politician,” which changes the subjective tone to a more objective one. 

These works are able to analyze biases in pre-trained language models and vision models individually, however, only a few works have been done on a multi-modal setting \cite{srinivasan_worst_2022}. In terms of these works, \cite{srinivasan_worst_2022} does research to demonstrate gender bias in VL-BERT, which works in a multi-modal setting. Additionally, \cite{thomas_predicting_nodate} collects a dataset of over one million unique images and associated news articles from left- and right-leaning news sources to develop a method to predict the image’s political leaning. \cite{thomas_preserving_2020} models a real-world scenario where image-text pairs convey complementary information with little overlap. The approach used in \cite{thomas_preserving_2020} helps preserve the semantic relationships between paired images and paired text. We can generate a series of semantically aligned images from \cite{thomas_preserving_2020} given a text or news headline. 

The objective of our project is to remove the subjectivity of the news and replace the news images with semantically aligned images that are politically less biased. Using mulitmodal information to reduce bias allows models to leverage a wider amount of information and reduce bias more effectively than uni-modal models.
\subsection{Political Bias Identification (Images)}
Regarding work related to political image debiasing, some existing papers focused on finding and predicting political bias within images. For example, \cite{thomas_predicting_nodate} utilizes a two-step process where a model first learns relevant visual concepts of an image to enable bias prediction. Then, a visual classifier is trained upon that model. In terms of political image debiasing, \cite{hirota_quantifying_2022} aims to evaluate and proposes metrics for measuring bias and bias augmentation in image text pairs, as well as how current models and datasets perpetuate and increase bias, but does not provide work on how to remove bias from image and text caption pairs. Additionally, \cite{boxell_slanted_2021} studies non-verbal bias indicators in facial features of political images and statistical analysis of news articles published around the 2016 election. Other works, such as \cite{peng_same_2018}, use facial feature extraction methods to identify and correlate certain visual elements with various kinds of bias. This work shows that different political figures are displayed quantifiably differently depending on political orientation and outlet. Finally, several approaches to other more general image bias-related tasks could be extended to political bias identification. For identifying bias in images, some work focuses on reducing bias to attribute/label data provided with or extracted from the data \cite{fabbrizzi_survey_2022}. Other common approaches focus on learning from lower-dimensional representations or determining bias by cross-analyzing multiple datasets \cite{fabbrizzi_survey_2022}. A more recent and uncommon approach utilizes an ensemble classification system by training a low-capacity network and a high-capacity network to reduce bias \cite{fabbrizzi_survey_2022,clark_learning_2020}. Overall, while several methods allow for identifying bias in images, most of these methods are, to our knowledge, not applied or extended to the political domain. Moreover, some approaches are not viable to perform regardless since annotating and performing cross-analysis on multiple datasets can be far too computationally expensive.


\subsection{Political Bias Identification (Text)}
Multiple techniques have been used in previous works to identify and reduce bias in the text. \cite{pryzant_automatically_2019} and \cite{sinha_determining_2021} use pre-trained BERT transformers to locate and correct biased words. \cite{xia_demoting_2020} and \cite{odbal_examining_2022} use adversarial training to create a classifier to detect and correct hate speech. \cite{gangula_detecting_2019} uses an attention-based mechanism on the article headline network to detect bias in the article body to mirror the order in which humans read articles. \cite{chen_detecting_2020} uses a Gaussian mixture model to observe probability distributions of the frequency, positions, and order of information to detect article-level bias. Regarding text dataset annotation, \cite{dinan_multi-dimensional_2020} proposes a framework to decompose gender biases across multiple dimensions, including the gender of the speaker, the person being spoken to, and the person being talked about. 

However, these previous approaches don’t leverage or interact with available visual data or only focus on correcting specific types of bias, i.e., gender, politics, and hate speech. By also processing images available in the article, our approach can use the images to inform the text bias and vice versa, as well as substitute the biased images with semantically similar ones but evaluated with a lower amount of bias.

\subsection{Evaluations/Metrics/Losses in Similar Approaches}
Due to the varied approaches in bias detection and debiasing in related works, several different methods are also used for evaluations. Frequently, individual datasets are scraped and developed by separate groups for specific projects. This approach tends to result in datasets where articles, usually just represented through the text modality, have been manually labeled with one particular kind of bias \cite{chen_detecting_2020,gangula_detecting_2019}, lending the ability to use these labels in a supervised classification task which identifies an article or sentence as biased. Bias detection here can vary between a binary classification (biased vs. neutral) or multi-class (biased towards one of a set of groups). Other works use existing corpora to identify and potentially correct biased language \cite{pryzant_automatically_2019,wang_multibias-mitigated_2022}. In most cases concerning bias detection with these datasets, a popular evaluation metric is to check the model's accuracy in classifying it as biased in binary or multi-class settings.

When evaluating text debiasing methods, basic natural language metrics such as BLEU can be used \cite{pryzant_automatically_2019}. Human evaluations are also crucial for understanding the effectiveness of debiasing models, with graders manually deciding on how fluent an original and corresponding debiased text are and whether they have the same meaning \cite{pryzant_automatically_2019}.

\subsection{Politics Dataset}
We will use the dataset collected in \cite{thomas_predicting_nodate}. This dataset is collected from biased news sources (from left/right) on 20 politically contentious issues such as Abortion, Black Lives Matter, LGBT, Welfare, etc. This dataset has around 1.8 million images/articles in total. This paper uses crowdsourced annotations to annotate over 14,000 sets that contain bias labels for images and image-text pairs alongside other metadata. Various curation mechanisms have been used to clean up the data from news sources. Also, the crowdsourced annotations are curated with quality control. Although \cite{thomas_predicting_nodate} identifies the bias as left/right-leaning, this paper does not have labels for neutral news sources. 

Another dataset that we will use is the Wiki Neutrality Corpus (WNC). WNC contains 180,000 sentence pairs with biased and neutralized information, metadata, and additional contextual sentences, which were all scraped from Wikipedia edits that ensured texts were as unbiased as possible \cite{pryzant_automatically_2019}. This dataset will neutralize the biases in the text of political news articles.

\section{Proposed Approach}
\label{sec:proposed_approach}

\begin{figure}[h]
\centering
\includegraphics[scale=0.6]{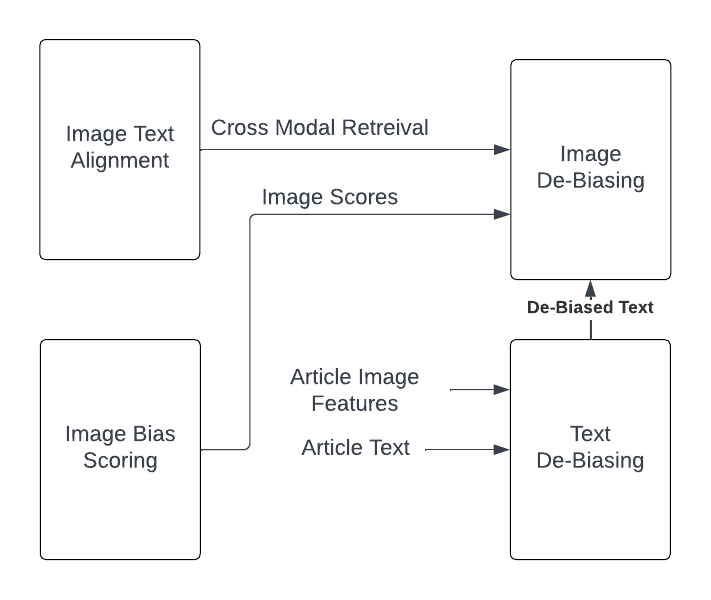}
\caption{Overall Architecture}
\label{fig:architecture}
\end{figure}
Figure \ref{fig:architecture} illustrates our overall architecture. 
Images and Texts that are used to train this model will be retrieved from various left-leaning, and right-leaning websites from the Politics Dataset \cite{thomas_predicting_nodate}. Based on their political leaning, we labeled the website from a score of -1 to 1 on a continuous scale. We scored -1 for far-left websites and 1 for right-leaning websites. We shared a score of 0 for neutral websites. For example, a score of -0.49 would be moderately left-leaning. We got these websites' scores from mediabiasfactcheck.com, from which the political dataset was used to determine left and right-leaning sources. For websites we did not have a score, we looked at their wording, sourcing, story choices, and political affiliation to give them a score from -1 to 1.

Section 3.1 will identify biased words in the articles using the BERT \cite{pryzant_automatically_2019} Biased Word Predictor. Then, we will train a BERT model with image and text inputs. Specifically, we will use Wikipedia images and text to generate an embedding space with unbiased images and text. During inference, we will mask the biased words and replace them with objective alternatives. 

Section 3.2 will use the CLIP \cite{radford_learning_2021} model to generate Cross-Modal Embedding Space with different biases. During inference, we will use the de-biased text from Section 3.1 to generate a de-biased image for our news article. 

Lastly, we will use the pre-trained Vision Transformer \cite{dosovitskiy_image_2021-1} used in Section 3.2 to predict the bias of each image in Section 3.3. We will fine-tune the Vision Transformer with more images and its respective labels. 
 
\begin{figure*}[h]
\centering
\includegraphics[width=\textwidth,scale=0.25]{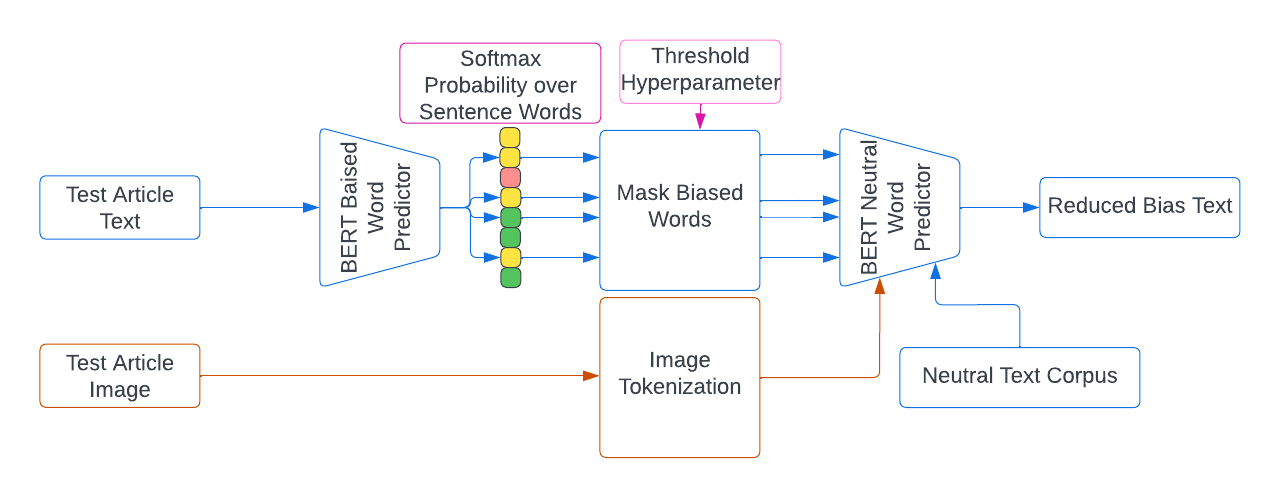}
\caption{Text Bias Neutralization}
\label{fig:text_bias_neutralization}
\end{figure*}

\subsection{Text Bias Identification and Neutralization}
Our architecture for Text Bias Identification and Neutralization can be seen in Figure \ref{fig:text_bias_neutralization}. First, we use the detection module from \cite{pryzant_automatically_2019} to find biased words from news articles. Given an input sentence, we will generate an output sentence that is semantically similar to the input, but the bias will be neutralized. The source sentences will be passed through a BERT model \cite{devlin_bert_2019} to determine the bias probability of each word of the source sequence using the technique presented in \cite{pryzant_automatically_2019}. The source sequence will be from the Politics Dataset's news article.

We will fine-tune a visual BERT model with paired neutral image/text data (Wikipedia images and text) \cite{li_visualbert_2019}. First, we will use the biased words detected from the detection module and mask them. After that, we will encode the article images into a sequence of tokens. Finally, the BERT Neutral Word Predictor will take in the encoded image sequence and the masked article text to predict the masked biased words and create a new sentence that has reduced bias.

\begin{figure*}[h]
\centering
\includegraphics[width=\textwidth,scale=0.2]{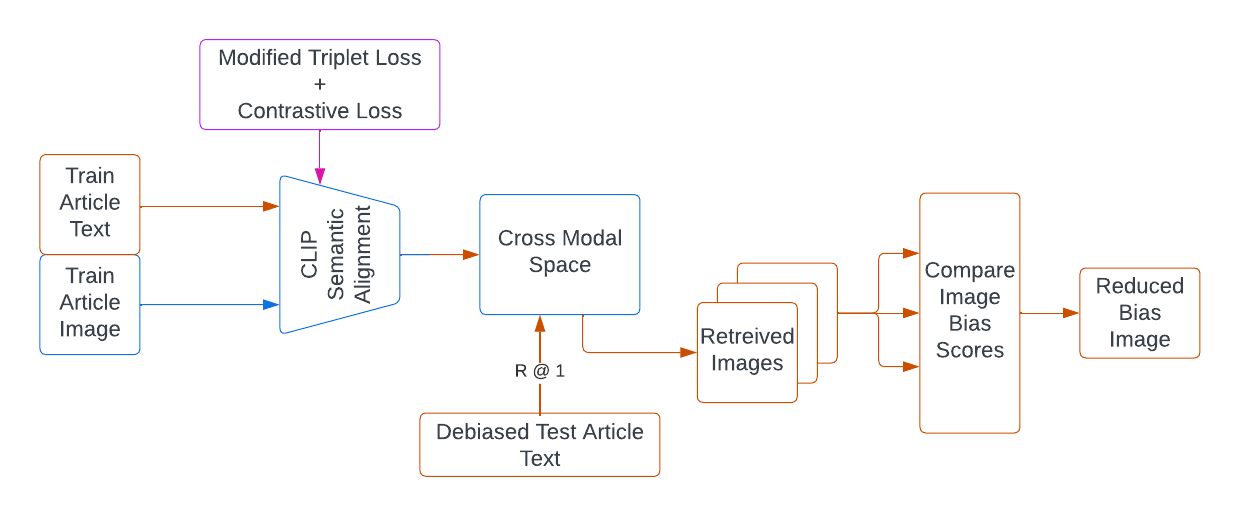}
\caption{Semantic Alignment Model}
\end{figure*}

\subsection{Semantically Aligned Images}
\label{sec:aligned_images}
The second primary step of the approach is performing semantic alignment based on the bias score introduced in the description of the overall architecture. We would do this by creating semantic neighborhoods following the process of \cite{thomas_preserving_2020}. Angular loss, an alternative to triplet loss, is used with a CLIP model to develop semantic alignment between images using existing semantics in the text \cite{wang_deep_2017, thomas_preserving_2020}. The angular loss is:
\begin{equation}
    \label{eq:ang_loss}
    L_{ang} = [||x_a-x_p||^2-4 \tan^2\alpha||x_n-x_c||^2]
\end{equation}
 Here, $x_a$ represents an anchor image, $x_p$ is a semantic nearest neighbor of $x_a$ in the Doc2Vec space, $x_n$ is a random negative sample where that is not $x_a$, and $x_c=\frac{x_a+x_p}{2}$ \cite{thomas_preserving_2020}. A semantic nearest neighbor is one that is one of the 200 nearest neighbors to the anchor in Doc2Vec space. This loss can also be applied to the text representations of each sample to keep semantically related text close in the embedding space. This loss enforces that the texts are semantically placed close to each other and using them to create a link between their respective images so that visually distinct images with similar meanings will be nearby in the semantic space. The ground-truth semantic similarity would be generated using a pre-trained Doc2Vec model because an unsupervised embedding learning was needed to gauge original semantic similarity, and Doc2Vec can do this while providing relative semantic representations between samples.
 
\begin{figure}[h]
\centering
\includegraphics[scale=0.8]{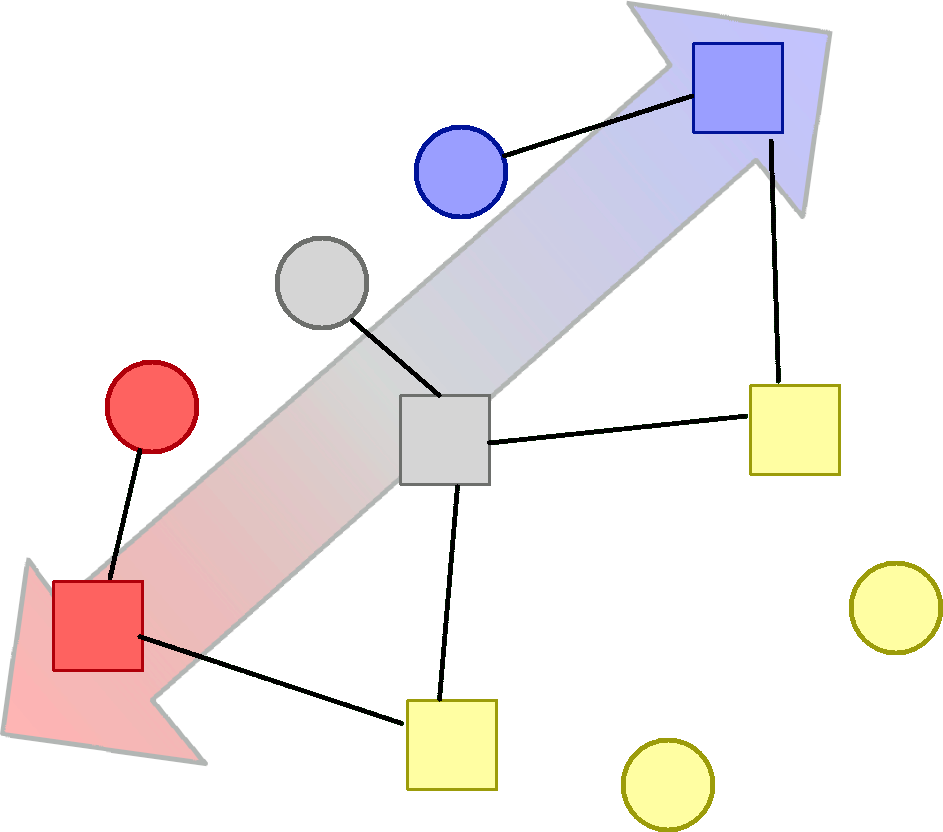}
\caption{Bias Loss. Text embeddings are represented as circles and image embeddings by squares. Here, the red, gray, and blue, embeddings have already been pulled together by semantic alignment loss. The bias loss then aligns the image representations in space based on their bias. Images that have a similar bias are pulled away from one another while being pushed away from images with different bias, creating regions of bias scores in the embeddings space. Debiased text that is fed into the space for retrieval should be near images that are semantically similar and similar (neutral) in bias.}
\label{fig:image_debias}
\end{figure}

 A new loss specific to this image debiasing task is also developed to account for the bias of images in this shared space. Aside from the angular loss that draws together distinct images of the same topic, the images should also be separated by bias. This process is done with another angular loss objective in which we create positive pairs using an image with similar bias of the anchor and negative pairs using an image with a different bias score. Again, this can be formulated as:
 \begin{equation}
    \label{eq:bias_loss}
    L_{ang} = [||x_a-B(x_a)||^2-4 \tan^2\alpha||x_n-x_c||^2]
\end{equation}
 where $B_b(x_a)$ is a function that randomly samples an image from the bias neighborhood of $x_a$. Images within a bias score range of 10\% of an anchor image are considered as part of the anchor's bias neighborhood. The range of 10\% can be changed, but 10\% was the starting range because our manual scoring typically varied by 5 or 10\% of other scores. This way, the original angular loss will group semantically-aligned images, and the images will be further grouped based on bias scores (e.g., the semantic-alignment loss provides a group of images about the topic “climate change” and the new bias loss separates those images into a spread of left, right, and neutrally biased scores). Figure \ref{fig:image_debias} visualizes how this loss affects the embedding space. With this, retrieval can be performed with debiased text, which will locate the nearest image that is both semantically similar and neutrally biased. Here, the model would take the debiased text as the input and feed it through a text-to-image cross-modal retrieval process to retrieve an image as the output. The nearest semantic image would be the designated debiased image allowing the model to perform image debiasing by “replacing” the query image with a less biased alternative with similar semantics. We will use the Image Bias Prediction in Section 3.3 to get the least biased image if the manually-scored bias is not already available in the database. This score-based system was used in place of a classification-focused approach because we wanted to capture a more precise granularity of bias which we can do using a scale rather than classes.

\begin{figure}[h]
\centering
\includegraphics[scale=0.8]{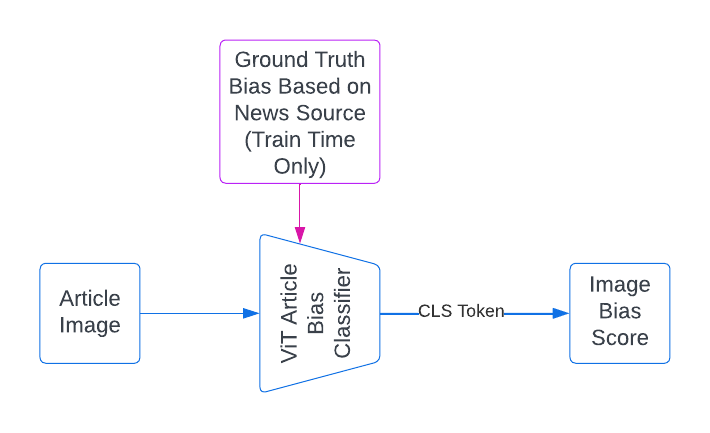}
\caption{Image Bias Prediction}
\label{fig:image_bias_prediction}
\end{figure}

\subsection{Image Bias Prediction}
The ViT model in Figure \ref{fig:image_bias_prediction} is trained in Section \ref{sec:aligned_images}. We will fine-tune this model further with more images and its respective bias score. Finally, we will pass an image through this model at inference, giving us a score from -1 to 1.


\section{Results}

\subsection{Bias Word Identification}
\begin{figure}[h]
\centering
\includegraphics[scale=0.3]{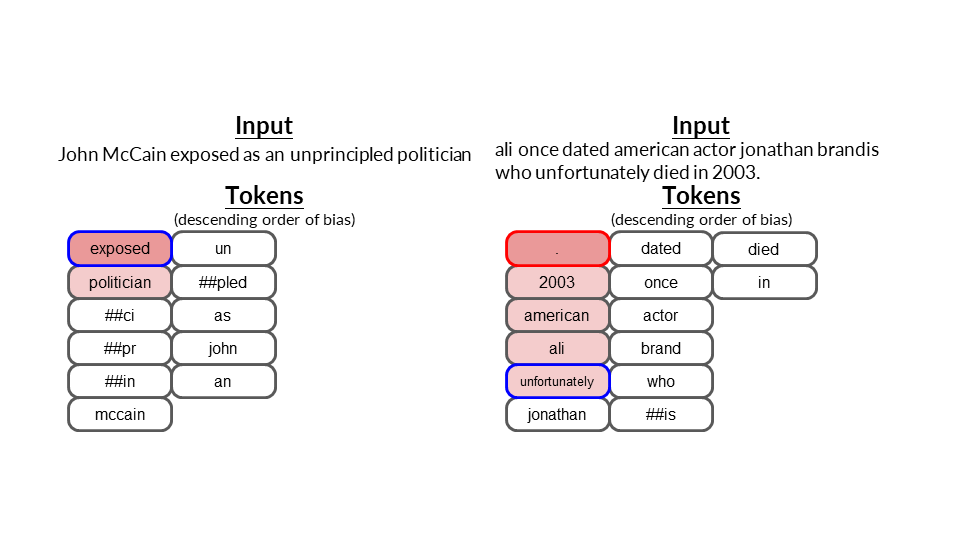}
\caption{Predicting the bias of BERT tokens with examples from \cite{pryzant_automatically_2019} and WNC, respectively.  The ground truth value is highlighted with a blue border.  The words with a higher probability for bias ($>0.5$) are shaded light red, with the most biased being shaded a darker red.}
\label{fig:bias_identification_1}
\end{figure}

\begin{figure}[h]
\centering
\includegraphics[scale=0.3]{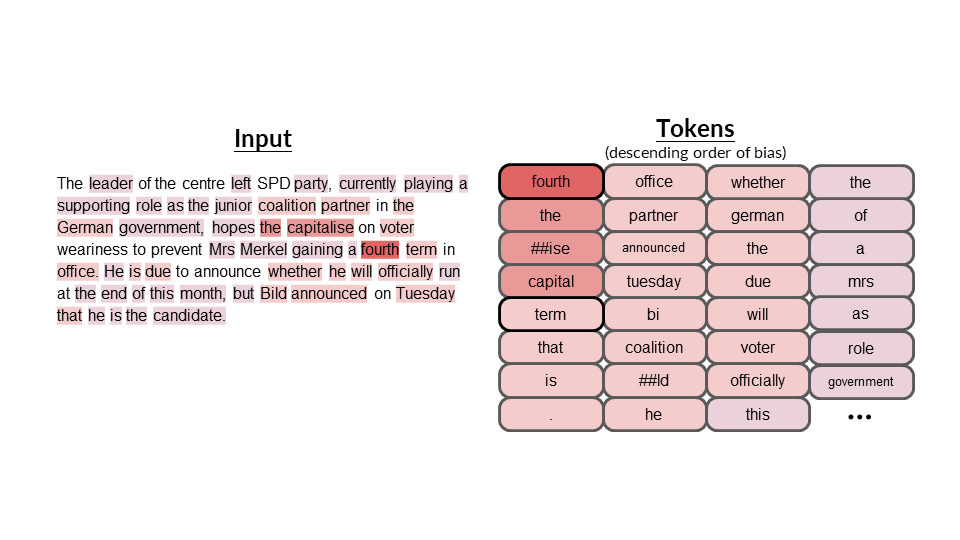}
\caption{Predicting the bias of BERT tokens with examples from the Politics Dataset.  The color is: red for most biased, lighter red for $> 0.9$, pink for $>0.75$, light purple for $>0.5$, and no color otherwise.  The corresponding words in the text are highlighted with this color scheme.}
\label{fig:bias_identification_2}
\end{figure}

For the bias word identification algorithm, quantitative evaluations were considered but the quantitative results from \cite{pryzant_automatically_2019} were generated in conjunction with the neutralization portion of the algorithm, so there was no appropriate baseline to consider.  Furthermore, it is difficult to perform quantitative results on our own dataset because we do not have a ground truth component unlike the Wikipedia Neutrality Corpus (WNC), which contains debiased forms of biased text.  For these reasons, a qualitative approach is considered instead which can be seen in Figure \ref{fig:bias_identification_1}, and 
Figure \ref{fig:bias_identification_2}.  For implementation, the hidden size is 768, the number of layers is 12, and the learning rate for the results given is 0.0001.  Due to time constraints, the model was trained on at most one epoch of the dataset.

The results showcase that the detection model is often helpful in identifying the most biased word.  For the first example of Figure \ref{fig:bias_identification_1}, the exact word is identified, and for the second example it is within the top 5 biased words.  For Figure \ref{fig:bias_identification_2}, since there is no ground truth, potential improvements in regards to bias need to be visually identified.  For this example, "the [sic] capitalise" is the best candidate for change because the phrase implies that the author of the text believes that the leader in question is taking intentional advantage which is subjective.  All tokens associated with this phrase are within the top 5 and have a probability of $> 0.9$. 

However, this model does appear to be sensitive to the presence of punctuation, rare words, and long sentences.  In the second example of Figure \ref{fig:bias_identification_1}, there are several tokens placed ahead of the ground truth token, being ., 2003, american, and ali.  It is possible that the model may require more training to generate more probable results, as 2003, ali, and most punctuation do not imply any bias.  The model also degrades as the statements get longer, which may be expected because the training is performed on a dataset containing statements that are shorter on average compared to the ones on the Politics Dataset.  

\subsection{Bias Word Neutralization}
To evaluate the performance of the biased word neutralization, we evaluate the cosine similarity between the original word and the newly generated word. This is done because we want the new word to be semantically similar to the original word, so they should have similar vectors in a word vector embedding space. For the evaluation, the word vectors are generated from a pre-trained word vector trained on 16B tokens from Wikipedia, the UMBC web-base corpus, and statmt.org news datasets. The word vector model is available here \cite{noauthor_english_nodate}.

Over 1000 test samples from the politics dataset were first passed through the biased word identification module then used to evaluate the bias word neutralization module, The bias word neutralization module achieved an average cosine similarity of 0.3960 between the original word and the re-predicted word. We can compare this amount to the cosine similarity of a pair of synonyms, eg. the words "vacation" and "holiday" produce a cosine similarity score of 0.2400. The average of 0.3960 is substantially higher indicating that the model is not producing synonymous words in the majority of cases. Qualitatively we can see that in some cases it makes sensible predictions, however it fails in cases where larger context is needed. For example in a sentence talking about a military members discharge, the module replaced the masked word "discharged" with "graduate", clearly missing the broader context that the person was leaving the armed force rather than joining them.

\subsection{Image Alignment }
Additionally, a pre-trained CLIP model is used and fine-tuned using a modified triplet loss to improve semantic alignment between text and images in a shared embedding space \cite{thomas_preserving_2020,radford_learning_2021}. Performing this semantic alignment using text to guide the images provides the backbone for finding potential replacement images of the same object and possible images of concepts that relate to the image in context, i.e., connect to the same topic as that of the image. Image debiasing is a retrieval task, with heavily biased samples being processed and used to find images of similar objects and topics with less bias. To test its retrieval ability, the bias from the newly retrieved image will be measured and compared against the bias of the original image. Equation \ref{eq:sai_avg} measures the average bias of retrieved images in the test set, and Equation \ref{eq:sai_diff} measures the average divergence towards neutrality of the retrieved images as compared to the original images in the test set.
\begin{equation}
\label{eq:sai_avg}
\frac{1}{|Y|}\sum_{y \in Y}|b(N(y))|
\end{equation}
\begin{equation}
\label{eq:sai_diff}
\frac{1}{|Y|}\sum_{x,y \in X,Y}(|b(x)|-|b(N(y))|)
\end{equation}
where $X,Y$ is the set of test images and text, respectively, from the Politics dataset, $b$ is the bias function of an image that returns the ground truth bias for known images in the embedding space and estimated bias for newly scored input images, and $N$ is a function that returns the nearest image to the input text, $y$. A low average bias and high average difference would indicate that retrieved images are both neutral on an absolute scale and relative to the original images. 

\begin{figure}[h]
\centering
\includegraphics[scale=0.6]{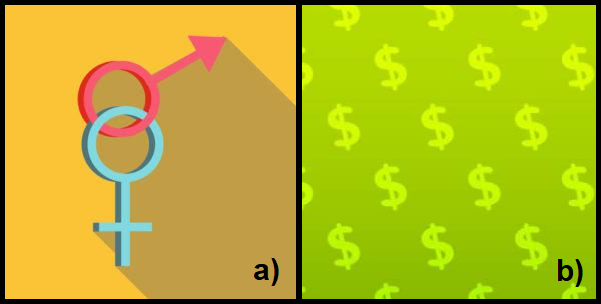}
\caption{This is an example of retrieved images from the testing space of the last image debiasing checkpoint. a) More biased input text "Politician has been exposed as anti-science." retrieved a somewhat generic image from the right-leaning, climate change category. b) More neutral input text "Politician has been described as anti-science." retrieved a different image from the right-leaning, minimum wage topic.}
\label{fig:img_debias_result}
\end{figure}

Qualitative evaluations are also important for evaluating retrieval effectiveness. Figure \ref{fig:img_debias_result} shows a pair of images retrieved from biased input text "Politician has been exposed as anti-science." and neutral text "Politician has been described as anti-science." These texts were not taken from the dataset as time constraints prevented final merging of all modules. The input text only deals with the topic of science, while the relation of each image has unclear implications in relation to science as well as unclear bias considering the topic. This could be due to the decreased training set size or the relatively low training time for the alignment. It is also possible that some aspect of the bias loss interferes with the semantic alignment in a negative way, which could indicate that more ablations with the bias weighting could be done.

\subsection{ViT model}
Our ViT model gives us a score of -1 to 1 during inference. Since our images are labeled from a score of -1 to 1, we can test the model's accuracy using two metrics, namely Root Mean Square Error (RMSE) and \(R^2\) score.

RMSE score gives us the standard deviation of the residuals or prediction errors. So, for example, if we have an RMSE score of 0.1, our model provides the results of a + or - 0.1 score close to the predicted output. A score of 0.1 is a good RMSE score since our model is predicting closely to the model's labeled value—however, a score of 0.5 shows that the model needs to predict the labeled score better.

\(R^2\) score, also known as the Regression Score Function, shows how well the model fits the data. We could have a negative \(R^2\) score, which indicates the model can be arbitrarily worse. A score of 0 shows that the model does not explain the variability of the response data around its mean. A score of 1 corresponds to a model that explains the variability of the response variable around its mean.

Due to lack of adequate resources, we were unable to get the \(R^2\)  and RMSE score. While training the model, we got encouraging results for our training and validation loss. 

\subsection{Human Evaluation}
While our report does not include any results from the human evaluation, a template form can be found here: (). The goal of this form is to verify that randomly selected de-biased text and image pairs still make semantic sense together, reduce the intial bias, and maintain the same semantic meaning as the original text and image pair.

\section{Conclusion}
Our attempt to de-bias both the text and images of our news succeeded as we devised an architecture that can de-bias emotionally charged language from the news. However, our results could have been more impressive because we needed more time and resources to train our model. Our model has four key components: Bias identification from text, Bias neutralization, Image Alignment, and Image Bias Score Prediction. We were able to identify the biased text for most of the cases correctly. However, the model is also sensitive to rare words and punctuation. Another model for bias neutralization has issues with understanding context since we only train one sentence at a time. For Image Alignment, the retrieval is computationally fast, but the retrieved results did not produce the desired results because of a lack of resources. Finally, we received good loss scores for the Image Bias Prediction Score for training and validation. Overall, our project was a step toward de-biasing text and images in the news.

\section{Ethical Considerations}
While the this approach is intended to reduce political bias in news that is supposed to cover objective truths, it is important to recognize that there are potential situations in which the technology could itself be biased or misused. Because bias scores must be manually assigned to each source in the dataset, there is potential for human bias to affect the grading of the data. In turn, this could negatively impact the downstream debiaser, for example if scorers tended to provide slightly more left-leaning scores, then a retrieved replacement image may be slightly more left-leaning than normal.

There are additional ethical considerations to examine as well. When would debiasing an article's images or text be considered censorship? In other words, how does one determine reasonable situations in which to use this debiaser or accept its results?
This work also only considers bias as a value on a single spectrum. Additionally, this work takes an America-centric view of political media which is likely different from the political climate of other countries. Issues could arise both from this simplistic single-spectrum view as well as from the differing ranges of political opinions across different regions. Depending on these political and societal differences between countries, certain media could seem more biased or neutral than it would somewhere else.

Another ethical question that arises is: if the negative or positive aspects being covered are truthful, is it appropriate to replace an image that reflects that emotional element with one that does not? For example, in an article about human rights abuse, would it be appropriate to replace an image of distressed people with one of people in a similar situation who do not seem distressed?
This is more difficult to separate because the truthful information being conveyed could be inherently evocative. Addressing this with a debiasing model could help remove a political slant to an extent, but would not help correct for other contextual information that was left out that creates bias in the piece.

An additional consideration is one related to peoples' perceptions of the model itself. If vulnerabilities or adversarial exploitations are found, we do not want the model to be used as a veneer for neutrality even when the underlying content is not unbiased nor objective. In a similar vein, the model could be used to find more biased image replacements, as there are not explicit restrictions placed on the bias of retrieved images.

\section{Acknowledgment and Future Work}
We want to thank Dr. Thomas for providing this opportunity to work on this research project and for the Multi-modal Vision class, which has immensely increased our skills to critique papers, think critically, and come up with novel ideas from the current literature on Multi-modal vision, which is moving forward at a rapid pace. All authors contributed equally to the project as we individually worked on different parts of the architecture to get favorable results and improvements. In addition, all authors contributed equally to the Project Proposal, Project Status Report, and Project Presentation and writing of the Final Report.

While working on this project, we experienced several challenges, such as a need for more training resources and experience in research and knowledge of various tools. Despite these challenges, our project is a novel project which attempts to de-bias both images and texts. Since our image de-biasing is a retrieval task, it is fast. Also, successfully implementing our paper would allow people to read factually accurate news. Despite these strengths, our project can reduce the bias of the text and images to make information less truthful. Since our models are interrelated, errors in one model may significantly compound mistakes in de-biasing. Finally, people watch audio and video sources for news these days, and we need to address this fundamental problem.

For Future work, we want to examine the directionality of bias in embedding space. Also, researching new loss functions will likely make regions for neutral representations. Finally, since we retrieve various images from our model, we need some metrics to measure the variety of images. Eventually, for audio and video news sources, we could work on a project where these sources are timestamped with their respective biases. 

\section{Individual Contributions}
We divided the work into four parts, one for each of the modules. Amun completed the work relating to the ViT for bias scoring. Chase completed the work with the CLIP model for building the multimodal space with our bias-aware angular loss. Xavier completed the work relating to the bias text identification and using the MODULAR design, and Cedric created the module for masking and replacing the biased words with neutral words. All of us worked in even parts to annotate the politics dataset with the bias scores from \href{https://mediabiasfactcheck.com/}{https://mediabiasfactcheck.com/}.

\section{GITHUB LINK}
\href{https://github.com/cedricb13579/NewsDebiasing}{https://github.com/cedricb13579/NewsDebiasing}

\end{document}